\documentstyle[12pt,aaspp4]{article}
\newcommand{\ifpp}[1]{#1}
\newcommand{\ifms}[1]{}

\ifpp{\input psfig}
\newcommand{\HI}{{\rm H{\sc i\,}}}
\newcommand{\HII}{{\rm H{\sc ii\,}}}
\newcommand{\HeI}{{\rm He{\sc i\,}}}
\newcommand{\HeII}{{\rm He{\sc ii\,}}}
\newcommand{\HeIII}{{\rm He{\sc iii\,}}}
\newcommand{\unit}[1]{\ifmmode {\rm\ #1\,} \else {$\rm #1$} \fi}
\newcommand{\angstrom}{\unit{\AA}}

\newcommand{\etal}{et al.\,}
\newcommand{\Halpha}{\unit{H\alpha\,}}
\begin{document}
\lefthead{Bowyer, Bergh\"ofer, \& Korpela}
\righthead{EUV Emission in Galaxy Clusters}

\title{EUV EMISSION IN ABELL 1795, ABELL 2199, AND THE COMA CLUSTER}

\author{Stuart Bowyer, Thomas W. Bergh\"ofer and Eric J. Korpela}

\affil{Space Sciences Laboratory, University of California, Berkeley, CA
  94720-7450, USA}

\authoremail{bowyer@ssl.berkeley.edu}

\begin{abstract}
  
  We report new EUV data on the cluster of galaxies Abell 1795.  These data were taken
  well away from a detector defect which could have compromised earlier
  results on this cluster. Our new observations confirm the validity of the
  original data set.  However, we find our results are strongly influenced by
  the variation of the telescope sensitivity over the field of view and upon
  the details of the subtraction of the EUV emission from the X-ray plasma.
  We investigate these effects using our new data and archival data on Abell 1795, Abell 2199 and the Coma cluster. When we use the
  appropriate correction factors, we find there is no evidence for
  any excess EUV emission in Abell 1795 or Abell 2199. However, we do find
  extended EUV emission in the Coma Cluster using our new analysis procedures,
  confirming that in at least this cluster some as yet unidentified process is operative.

\end{abstract}

\section{Introduction}

Extreme ultraviolet (EUV) emission in excess of that produced by the
well-studied X-ray emitting gas
in clusters of galaxies has been reported in five clusters of galaxies from 
observations with the Extreme Ultraviolet Explorer (EUVE).  
The effective bandpass of the EUVE telescope employed in these observations
is defined by the intrinsic response of the telescope combined with the 
absorption of the intervening Galactic interstellar medium (ISM).  This
bandpass has a peak at 80 \AA\ with 10 percent transmission at 66 and 100 \AA.
A
variety of instrumental effects that might have explained these results have
been advanced but a detailed analysis has shown these factors cannot explain
the data (Bowyer, Lieu, \& Mittaz l998). A number of ISM effects
have been suggested which might have produced the EUV excess.  An error in the
measurement of the total amount of absorbing gas in the Galaxy could have
explained these results.  However, detailed measurements of the Galactic
neutral hydrogen in the direction of these clusters (Lieu et al., 1996a,b)
have eliminated this as an explanation.  A particular ionization state of the
Galactic ISM could have produced this result but a detailed analysis shows
that the required ionization is not, in fact, realized (Bowyer,
Lampton, \& Lieu 1996).

It is interesting to note that the EUV excess is detected in some ROSAT
images.  However, the effect is sufficiently marginal that the ROSAT results
can almost be explained away through the use of particular combinations of
intervening Galactic ISM and its ionization state, and different cross
sections for absorption by hydrogen and helium (Arabadjis \& Bregman 1999).
The EUVE results, however, cannot be explained in this manner.  It is also
interesting to note that the EUV excess has been reported in every cluster
investigated with EUVE.

A number of suggestions have been made as to the source of this EUV emission.
Initial work focused on additional components of ``warm gas" ($\sim 10^6$\,K).
The problem with this suggestion is that gas at this temperature is near the
peak of the cooling curve and substantial energy is needed to supply the
energy radiated away.  One mechanism that can provide this energy is
gravitational condensation.  Cen \& Ostriker (1999) have suggested that a
pervasive warm intergalactic gas constitutes the majority of matter in the
Universe; as this gas coalesces onto clusters of galaxies, it could produce
the energy needed to sustain the EUV emitting gas.

Several authors (Hwang 1997; En{\ss}lin \& Biermann 1998) have suggested the
EUV flux in the Coma Cluster is inverse Compton (hereafter: IC) emission
produced by the population of electrons producing the radio emission
scattering against the 3$^{\circ}$\,K Black Body cosmic background.  However,
Bowyer \& Bergh\"ofer (1998) have shown that the existing population of radio
emitting cosmic ray electrons cannot be responsible for the EUV emission in
the Coma cluster, and some other population of cosmic rays will be required if
this mechanism is the source of the EUV emission in this cluster.  Lieu et al.
(l999a) have suggested that the Coma cluster contains a large population of
cosmic rays which are producing the 25 to 80 keV emission seen by BeppoSAX
(Fusco-Femiano et al. 1999) and RXTE (Rephaeli, Gruber, \&
Blanco 1999) via IC emission . They propose this population of
cosmic rays extrapolated to lower energies will produce the observed EUV flux
by IC emission. However, these authors have not addressed the fact that this
population of electrons will produce a spatial distribution of the EUV flux
which is inconsistent with observational results (Bowyer \& Bergh\"ofer
l998).

En{\ss}lin, Lieu, \& Biermann (l999) have explored processes that might
produce a heretofore undetected population of lower energy cosmic rays which
could produce this flux. They demonstrate an evolutionary scenario in which
relativistic electrons produced in the last merger event in Coma two Gyrs ago
could produce these electrons.  However, this model cannot produce the spatial
profile of the EUV emission reported by Bowyer \& Bergh\"ofer. They also
consider IC scattering of starlight photons and show that in some scenarios
this could account for the EUV flux and the required spatial distribution.

Sarazin \& Lieu (1998) have suggested that all clusters of galaxies may
contain a relic population of cosmic ray electrons that are unobservable in
the radio and these will produce excess EUV flux by inverse Compton scattering
against the 3$^{\circ}$\,K cosmic background. Their proposal is based upon,
and explains, details of the EUV data obtained on Abell 1795 (Mittaz, Lieu \&
Lockman 1998).

Because of the key role of the EUV data from Abell 1795, we have obtained new
observations of this cluster with EUVE.  A new observation is especially
important because of peculiarities in the EUVE Deep Survey (hereafter DS)
telescope that may have affected the Mittaz et al. (1998) data set.  A dead
spot exists at the bore sight of this instrument and this dead spot was
centered near the cluster core in the observations of Mittaz et al. It is
only a few pixels in size and would not be expected to affect the observations
of a diffuse source. However, the pulse height of the photons detected in the
region surrounding this dead spot may be degraded in a manner that is not
known (Vallerga \& Roberts 1997). This effect, if present and unaccounted
for, could lead to a substantial loss of counts in the central region of the
image. We have taken special care to insure that our new observations of Abell
1795 were taken in a manor that was free from any spurious instrumentation
effects.

We find that the results obtained are crucially dependent upon the
characterization of the DS telescope, and upon details of the estimation of
the EUV emission from the X-ray plasma. We have examined
these aspects in depth. The results we obtain are quite different from those
obtained in previous work, and consequently we expanded our inquiry by examining
archival data on Abell 1795, Abell 2199, and the Coma cluster. We report the results of our
investigations herein. Throughout this paper we use H$_0$ =
50\,km\,s$^{-1}$\,Mpc$^{-1}$.

\section{Data and Data Analysis}  

Ninety thousand seconds of data on Abell 1795 were obtained with the DS telescope of
EUVE (Bowyer \& Malina 1991) in March of 1998.  The observations were taken
in two segments of 45,000 seconds each. The individual observations were obtained
13\arcmin\ to the east and to the west of the core of Abell 1795.  These
locations are well away from the central region in the detector that could
potentially have produced spurious results, and are near the point of optimum
focus for the DS Telescope.

The data were processed using procedures of the IRAF EUV package provided by
the Center for EUV Astrophysics (CEA, Berkeley) which were especially designed
for the analysis of EUVE data. As part of this process, we excluded detector
events with pulse heights far from Gaussian peak of the photon pulse-height
spectrum. Low energy events due to spurious detector noise($\approx$15\% of
the total), and high energy counts due to cosmic rays and charged particles
($\approx$25\% of the total), were screened out. A detailed description of
different background contributions to the DS data can be found in Bergh\"ofer
et al. (1998). We point out that the location of the Gaussian peak in the
pulse-height spectrum is not constant for all EUVE DS observations since the
gain of the DS detector was changed periodically in the course of the mission.
Consequently, pulse height limits were chosen individually for each DS
observation.The resulting filtered event lists were corrected for electronic
deadtime and telemetry throughput effects.

The background of the DS telescope consists of a uniform detector background,
$B_{\rm int}$, and a component that may vary over the field because of a
variety of effects including vignetting, variations in the thickness of the
filter covering the detector face, variations in the quantum efficiency over
the face of the detector, and other causes. Hereafter, we call this second
component the vignetted background, $B_{\rm vig}$. To investigate the
possibility of a field variation effect, we chose four 20,000 second observations
of blank sky with low and similar backgrounds that were obtained in a search
for EUV emission from nearby pulsars (Korpela \& Bowyer 1998). We added
90,000 second of data from a blank field at R.\,A.$_{2000} = 3^h31^m39^s$ ,
Dec.$_{2000} = +18^{\circ}28\arcmin33\arcsec$ obtained from the EUV
archives. We processed the data as described above. We established that once
proper pulse height selection of the detector events had been made, the
detector backgrounds were all spatially identical.

\ifpp{
\begin{figure}[tbp]
\begin{center}
\ \psfig{file=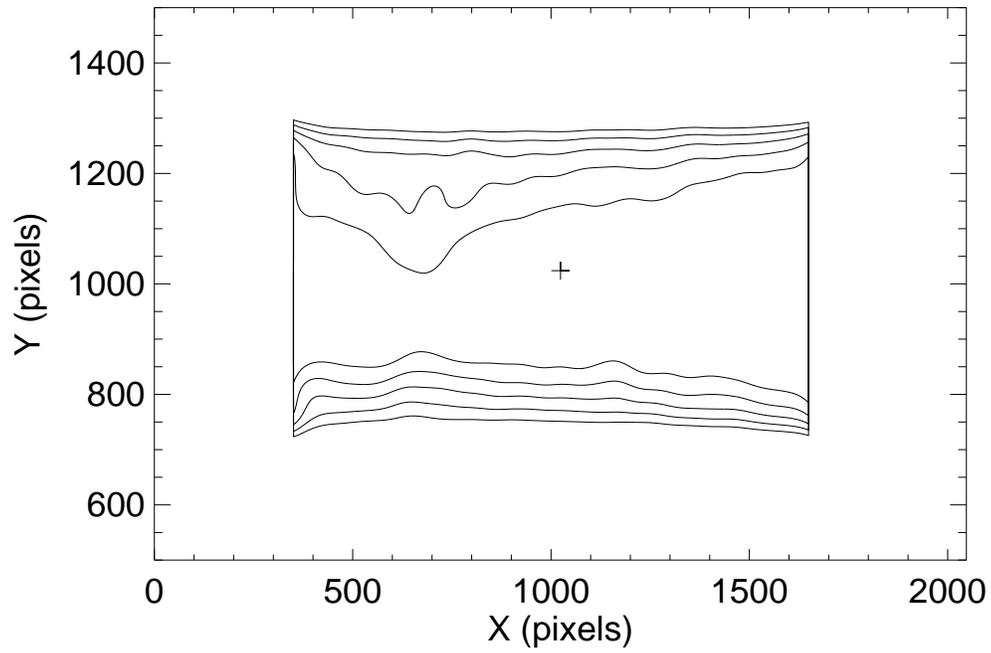,width=5.5in}
\caption{\small A contour plot of counts obtained in long duration DS
exposures showing the sensitivity variation of the DS Telescope over the
field of view. We have cut the regions at the detector ends where
detector distortions become severe. The field displayed is approximately
1.75 degrees x 0.73 degrees.}
\end{center}
\end{figure}
}

A contour plot of normalized count rates in these exposures convolved with a
32 pixel wide Gaussian is shown in Figure\ 1. Each contour represents a 10\%
change in the measured count rates.  \footnote{Investigators interested in
  using this observationally derived vignetted background may access Fits Files at
  ``http://sag-www.ssl.berkeley.edu/$\sim$korpela/euve\_eff'' .}  It is
informative to compare this observationally derived result with the
theoretically derived product provided by Sirk et al. (1997), which has been
used in previous work on this (and all other) clusters.

All observations contain both $B_{\rm int}$ and $B_{\rm vig}$. Because the
ratio of these two backgrounds can vary, we must correct for this effect when
scaling previously measured backgrounds to the backgrounds of our
observations. Our background subtracted image is:

\begin{equation}
I_{\rm net} = I_{\rm on} - B_{\rm int} - f B_{\rm vig} 
\end{equation}

where $I_{\rm on}$ is the on-source image.  The term $B_{\rm int}$ is derived
from measurements of the background in obscured regions of the detector
covering about 3.5 \% of the detector area.  The term $B_{\rm vig}$ represents
the vignetted background.  The factor $f$ is used to fit the vignetted
background levels in the blank field with those of the on-source image.  This
factor is derived by fitting the observed photonic background with that of the
blank field images in a region far from the source, $R > 15\arcmin$ in the
case of Abell 1795 and Abell 2199, and $R > 17\arcmin$ in the case of the Coma
cluster.  Because of the long duration of the background exposures, the
statistical errors in $f$ are less than 1\%.  When comparing on-source and
background in small detector regions, our errors are dominated by the count
statistics of the region, rather than errors in the background fitting.

In our observations of Abell 1795, the central source in the two offset exposures
were made at regions with similar efficiencies, so these images could have
been directly added without affecting the intensities of the central image.
However, the backgrounds away from the cluster core will be affected
differently so the vignetting correction was applied separately to each image
before adding.

We examined two extragalactic sources detected near the cluster center: PGC
94626 about 7\arcmin\ to the southwest, and 134834.3+262207 about 22\arcmin\ 
to the southwest, for use in adding the two images.  Each of these sources is
a point source and we found each image to be spherically symmetric.  We summed
the data and fit the resulting profile with a Gaussian.  The overall fit was
good (reduced $\chi^2$ = 1.2) but the wings were not well fit far from the
centroid. Adding a second Gaussian to reflect the extended nature of the point
spread function improved the fit (reduced $\chi^2$ = 1.05).  These sources
provide an estimate of the point-spread function of the detector in the
off-axis region where our cluster data was obtained; the FWHP of the images
are $\approx$ 25\arcsec.

\ifpp{
\begin{figure}[tbp]
\begin{center}
\ \psfig{file=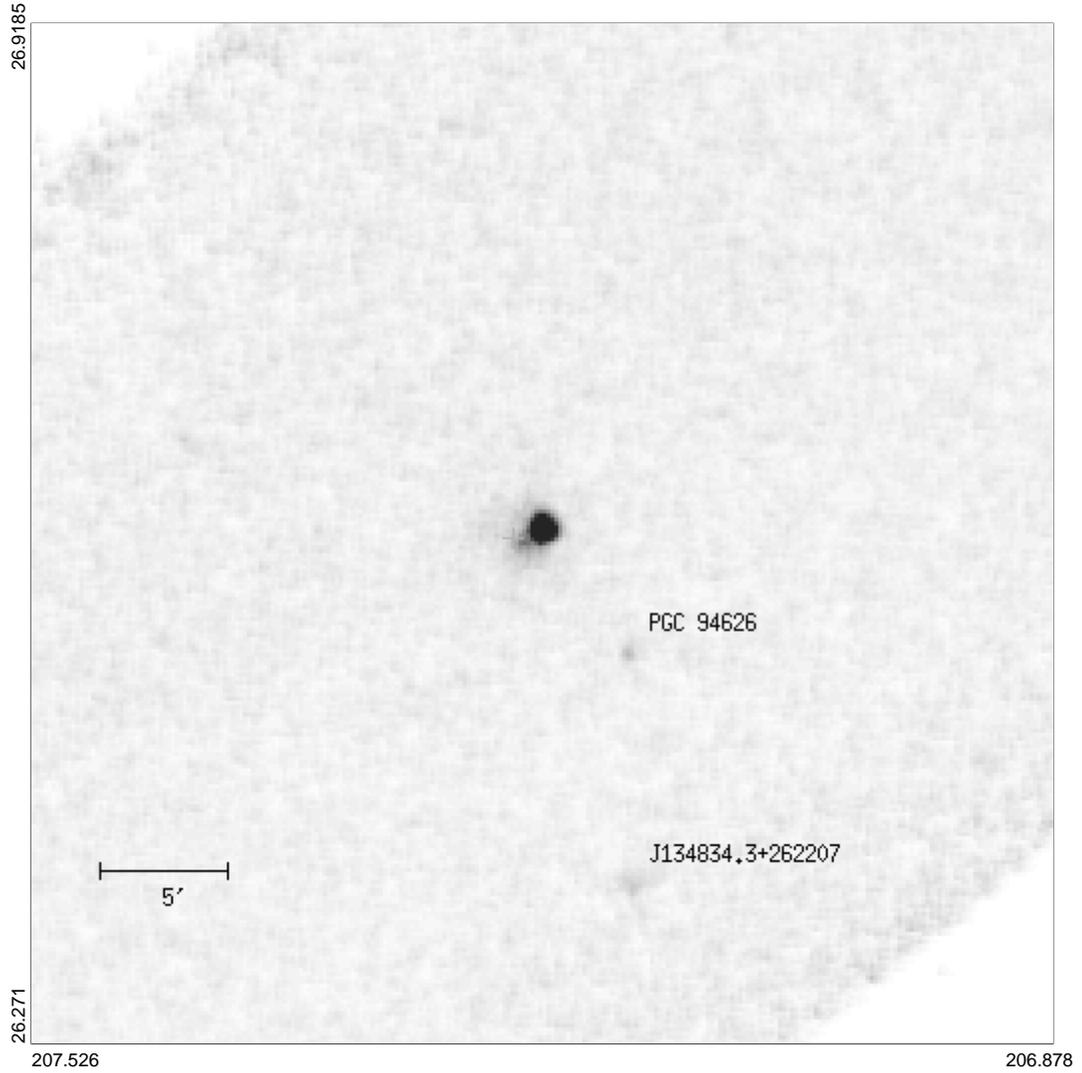,width=5.5in}
\caption{\small The spatial distribution of the EUV counts in Abell
  1795.  The zero points of the image are R.\,A.$_{2000} = 13^h48^m52^s$,
  Dec.$_{2000} = +26^{\circ}35\arcmin34\arcsec$. A bright EUV emitting
  transient is visually obvious near the cluster center.  The diffuse EUV
  cluster emission peaks at the position of the central galaxy in the
  cluster.}
\end{center}
\end{figure}
}

We added our two separate observations of Abell 1795 using the two
extragalactic sources as fiducials.  The resultant image is shown in Figure\ 
2.  A bright source is evident in this image northwest of the
center of the cluster.  Examination of other unrelated images showed no
detector flaws at this location and this source does not appear in the
previous observation of this cluster.  The radial profile of this emission
shows the source to be $\leq$ to the point spread function of the telescope
and is consistent with a point source. A time profile of the count rate
suggests the emission is the result of a transient source.
As improbable as this seems, we conclude a transient EUV
source appeared at this location in the sky at the time our observations were carried
out.

We have compared our new data on Abell 1795 with the archival data on this
cluster and find that our results at $R > 2\arcmin$ (which excludes the
effects of the bright point source in our new data set) are identical within
the counting errors, confirming the validity of the original data set used by
Mittaz et al. (1998). Because the two data sets are identical and the more
recent set is contaminated with a point source, we have used the archival data
on Abell 1795 for our subsequent analysis.

\ifpp{
\begin{figure}[tbp]
\begin{center}
\ \psfig{file=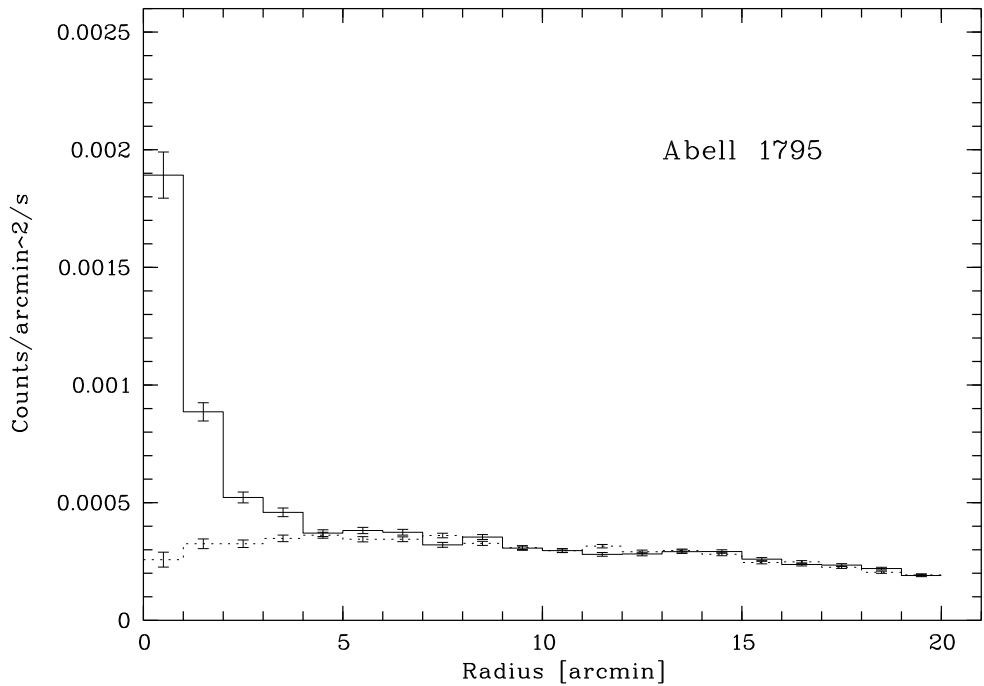,width=5.5in}
\caption{\small The azimuthally averaged radial intensity profile of
  the EUV flux in Abell 1795 is shown as a solid line. The vignetted
  background from long observations of blank fields is shown as a dotted line.
  There is no obvious excess EUV emission beyond 4\arcmin.}
\end{center}
\end{figure}
}

We derived the azimuthally averaged radial intensity profile of the EUV
emission of the cluster as a function of projected radius from the central
core assuming spherical symmetry. This profile is shown as a solid line in
Figure\ 3. Our vignetted background is shown as a dotted line. It is visually
apparent that there is no excess EUV emission at radii larger than 4\arcmin.
It is also clear that an improperly chosen background chosen at $R >
15\arcmin$ would result in apparent emission at smaller radii simply because
of the effects of the vignetted background in the DS telescope.

We next determined the expected intensity and distribution of the EUV emission
expected from the X-ray emitting plasma. We used the X-ray radial emission
profile provided by Briel \& Henry (1996). This profile was derived from ROSAT
PSPC observations of the cluster in the energy band between 0.5--2.4 keV. At
larger radii $(R > 4 \arcmin)$ this profile is well fit by a King (1972)
profile
with $\beta = 0.93$ and describes the large scale cluster X-ray
emission with a temperature of 6.7 keV. The ROSAT observations also show a
central excess emission within $R < 4\arcmin$.  Briel \& Henry (1996) obtained
a temperature of 2.9 keV for this excess.
We
derived conversion factors for counts in the 0.5--2.4 keV band of the ROSAT
PSPC to EUVE DS counts using these plasma temperatures.  Our
derivation employed the MEKAL plasma emission code with abundances of 0.3
solar. For a temperature of 6.7 keV we obtained a conversion factor of 126;
the value for 2.9 keV was 111. We found that varying the temperatures by
$\pm$ 1 keV or using different abundances only affect these conversion factors
by a few percent and thus  changes of this nature would not significantly alter our results.
We
found that a deprojection of the emission components which takes into account
the emission measures and sizes of the different components leads to the same
result.

The correction for the intervening absorption of the ISM in our Galaxy will
have a substantial impact on our results. Many workers simply apply the cross
sections of Morrison \& McCammon (1983) or Baluci\'nska-Church \& McCammon
(1992) for this correction, but there are several problems with this approach.
The \HeI absorption coefficient in
this work is incorrect (Arabadjis \& Bregman 1999). 
In addition, the ionization state of the ISM will substantially affect the
result.  The ISM absorption at EUV energies is primarily due to \HI, \HeI,
and \HeII; the metals in the list of Baluci\'nska-Church \& McCammon (1992) provide
less than 30\% of the absorption at wavelengths greater than 50 \angstrom, and
less than 10\% at wavelengths greater than 100 \angstrom, and none of the 
Galactic ISM is in the form of \HeIII (see discussion below).  
Hence in general terms the absorbing material and related factors are given by:
\begin{equation}
N({\rm H(tot)}) = N(\HI) + N(\HII)
\end{equation}
\begin{equation}
N(\HeI) = 
    \frac{1}{10}\left[ N({\rm H(tot)}) \right] \left( 1 - X(\HeII) \right)
\end{equation}
\begin{equation}
N(\HeII) = 
    \frac{1}{10}\left[ N({\rm H(tot)}) \right] \left( X(\HeII) \right) \\
\end{equation}

\ifpp{
\begin{figure}[tbp]
\begin{center}
\ \psfig{file=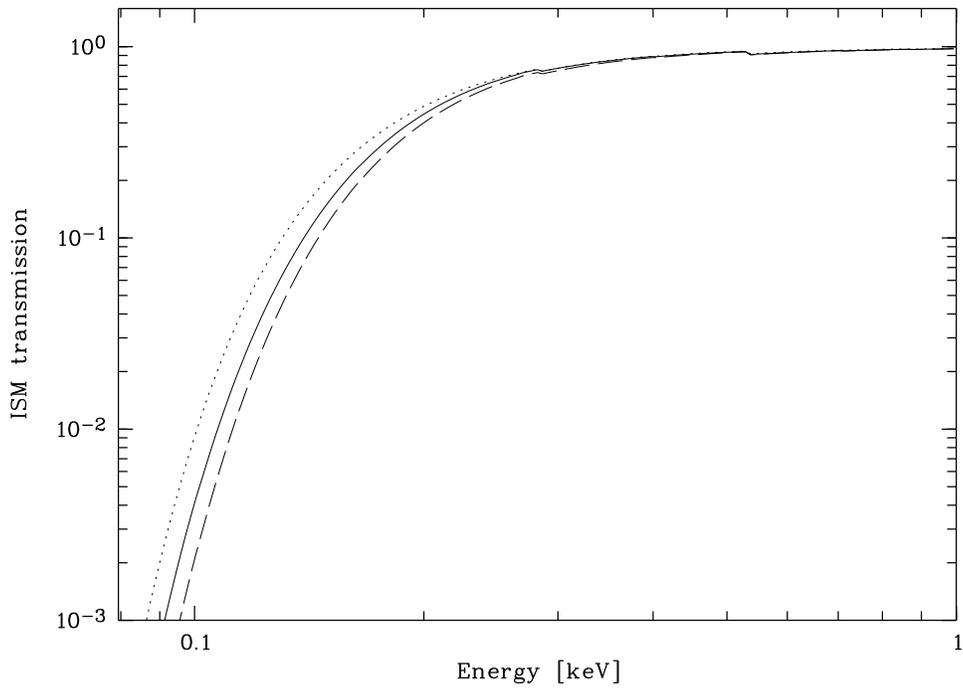,width=5.5in}
\caption{\small The effects of different ISM effective cross sections
  and assumptions as to the ionization state on the absorption of EUV emission
  . The absorption, normalized to one at 1 keV, is shown as a function of
  energy: Baluci\'nska-Church \& McCammon (1992),dashed line; Morrison \&
  McCammon (1983) dotted line; cross section used herein (see text), solid 
  line.}
\end{center}
\end{figure}
}

We have calculated the Galactic ISM absorption using these columns with \HI cross sections of Rumph \etal (1994), \HeI
cross sections from Yan \etal (1998), and \HeII cross sections from Rumph \etal
In Fig.~4, we compare the nominally applied cross sections of Balucinska-Church 
\& McCammon (1992) and Morrison \& McCammon (1983) with the cross section used
herein which includes the improved cross section for \HeI and the ionization
fractions as described.
We used these values with an improved estimate of the Galactic
neutral hydrogen column density in the direction of Abell 1795 of N(H{\sc I})
= $0.95 \times 10^{20}$cm$^{-2}$ (J. Lockman, private communication) .
We assume the total helium is 10\% of the total hydrogen column.
A direct measurement of the \HII column can be obtained, in principle, from
measurements of the \Halpha flux in this direction (Reynolds \etal, 1998).
Unfortunately, only an upper limit to this flux of $1 \times 10^{19} 
{\rm cm}^{-2}$ is currently available (Haffner, private communication).
A reasonable estimate for the \HII column, based on all the available data, is 
that it is close to this upper limit (Reynolds, private communication).
Consequently we have used this value for the \HII column.  The
amount of \HeII in this direction can be obtained from Fig.~1 of Bowyer
\etal (1996).  
For A1795, $N({\rm H(tot)}) = 1.1 N(\HI)$,
$N(\HeI) = 0.1 [1.1 N(\HI)](1-0.02)=0.108 N(\HI)$, and $N(\HeII) = 0.1 (1.1 N(\HI)] \times 0.02= 2.2\times 10^{-3} N(\HI)$.
\ifpp{
\begin{figure}[tbp]
\begin{center}
\ \psfig{file=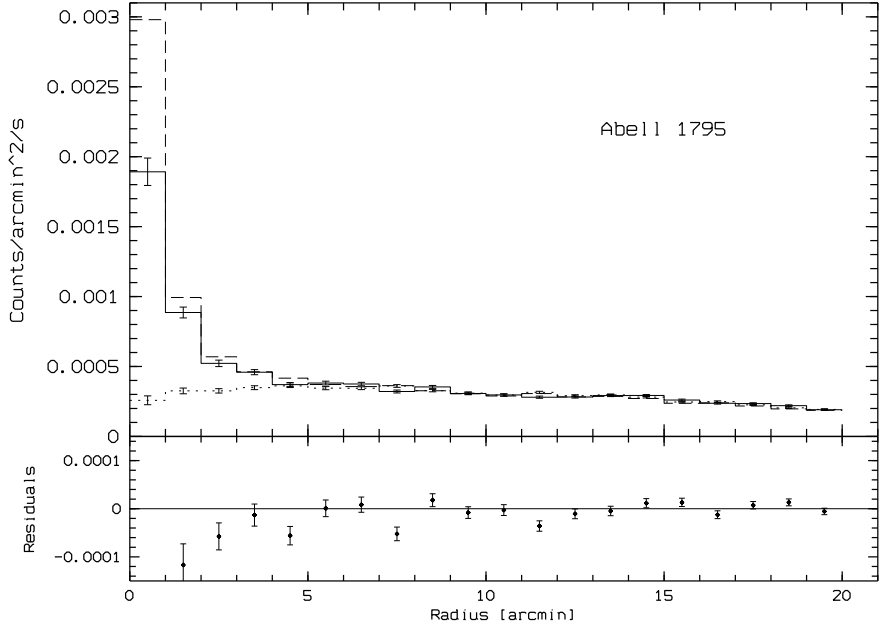,width=5.5in}
\caption{\small The expected EUV emission from the X-ray plasma in Abell
  1795 is shown as a dashed line. The azimuthally averaged radial intensity
  profile of the EUV flux is shown as a solid line. The vignetted background
  is shown as a dotted line.}
\end{center}
\end{figure}
}
The absorption corrected results for A1795 are shown in Figure\ 5. The 
observed EUV emission is {\it
  less} than that produced by the x-ray plasma. This appears to be unphysical but
is simply understood as discussed below.
\ifpp{
\begin{figure}[tbp]
\begin{center}
\ \psfig{file=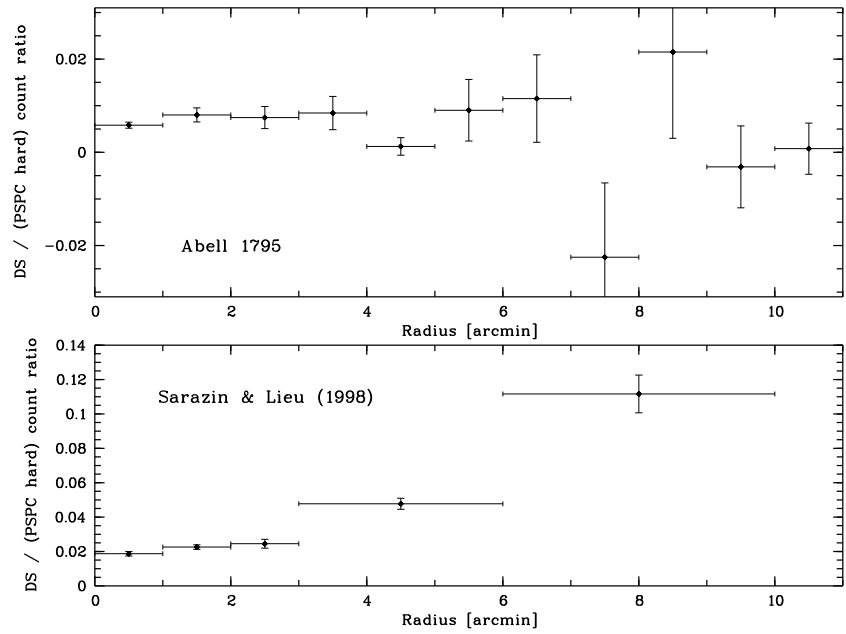,width=5.5in}
\caption{\small In 6a we show the ratio between our background
  subtracted EUV radial emission profile obtained with the EUVE Deep Survey telescope and the X-ray profile obtained with
  the ROSAT PSPC. In 6b we show the Deep Survey to PSPC count rate ratio as a function
  of radius obtained by Mittaz et al. (1998) and used by Sarazin \& Lieu
  (1998).}
\end{center}
\end{figure}
}

In Figure\ 6a we show the ratio between our background subtracted EUV radial
emission profile and the X-ray profile obtained with the ROSAT PSPC.  Within
the inner 4\arcmin\ this ratio is almost constant. At larger radii the ratio
is consistent with zero within the errors, confirming the EUV signal is absent
leaving only background noise. For comparison, Figure\ 6b shows the DS to PSPC
count rate ratio as a function of radius obtained by Mittaz et al. (1998) and
used by Sarazin \& Lieu (1998).

\ifpp{
\begin{figure}[tbp]
\begin{center}
\ \psfig{file=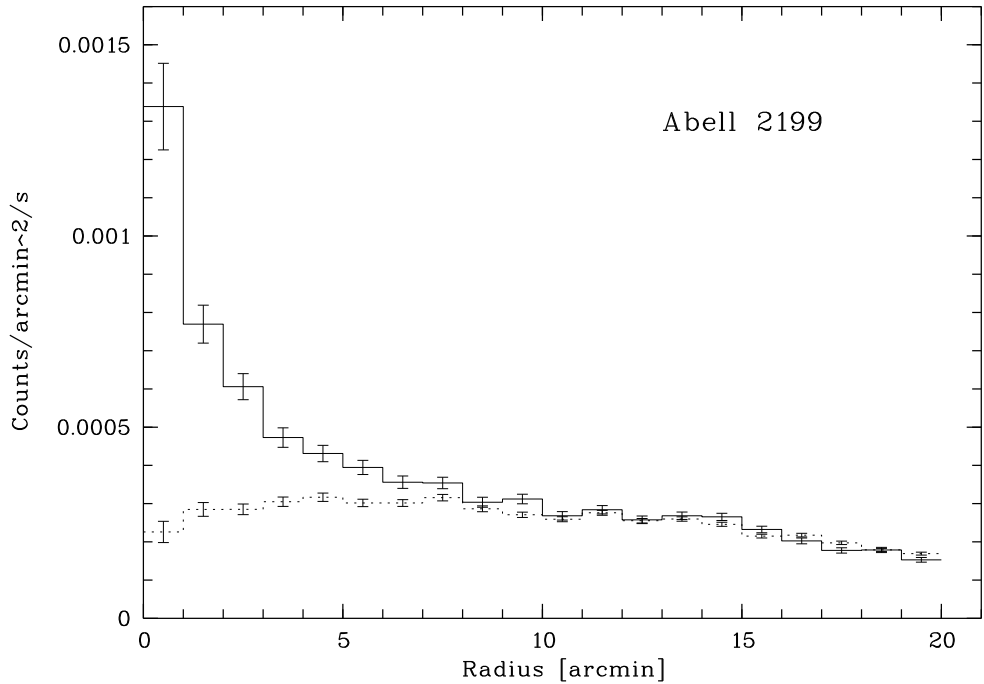,width=5.5in}
\caption{\small The azimuthally averaged radial intensity profile of the
  EUV emission in Abell 2199 is shown as a solid line. The dotted line is the
  vignetted background. There is no obvious EUV emission beyond 8\arcmin.}
\end{center}
\end{figure}
}

In view of the similarity between the distribution of the EUV emission
reported for Abell 1795 and Abell 2199 (Lieu, Bonamente, and Mittaz 1999b), we
examined archival data on Abell 2199 to ascertain whether a vignetted
background could have produced an artificial extended diffuse EUV halo in this
cluster. In Figure\ 7 we show the radial profile of the raw EUV data and the
vignetted background for Abell 2199. It is apparent that there is no excess EUV emission
beyond 8 \arcmin. We use the results of Siddiqui, Stewart, and Johnstone 
(1998) to model the EUV emission from the X-ray gas in the cluster.
They found T(core) = 2.9 keV and T(outer) = 4.08 keV.
The conversion of the ROSAT
X-ray count rates into EUVE DS count rates has been done as
described for Abell 1795. For Abell 2199 we found DS to PSPC hard band
count rate ratios of 83  for T = 2.9 keV and 89   for T = 4.08 keV.
Absorption by the Galactic ISM was accounted for using N(\HI) of $8.3 \times
10^{19} {\rm cm}^{-2}$ (Lieu, \etal 1999a) with ionization fractions and
cross sections as described previously.  The results are shown in Fig.~8
as a dashed line.
Again,
the expected EUV emission from the X-ray gas is larger than the observed flux.
\ifpp{
\begin{figure}[tbp]
\begin{center}
\ \psfig{file=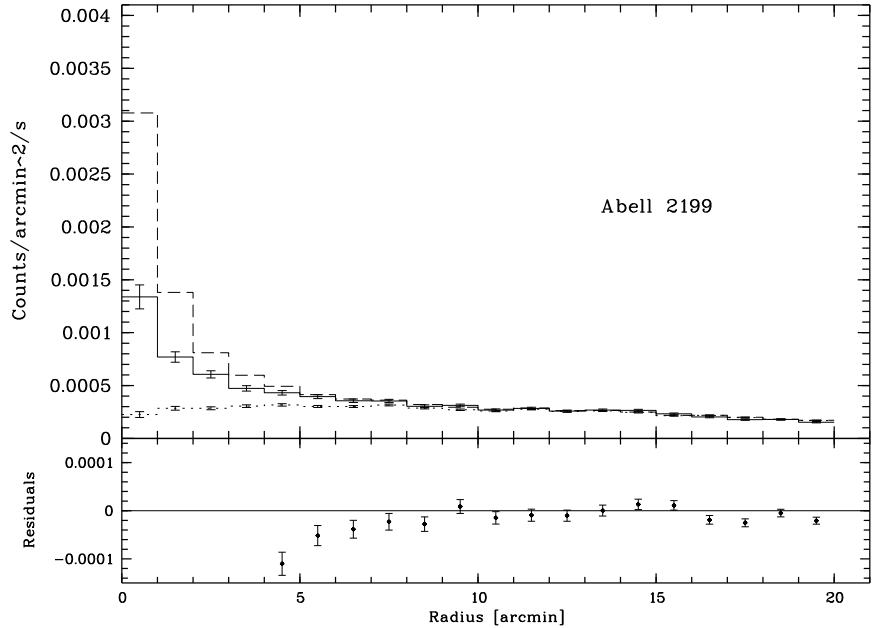,width=5.5in}
\caption{\small The expected EUV emission from the X-ray plasma in
  Abell 2199 is shown as a dashed line. The EUV flux is shown as a solid line.
  The vignetted background is shown as a dotted line.}
\end{center}
\end{figure}
}
\ifpp{
\begin{figure}[tbp]
\begin{center}
\ \psfig{file=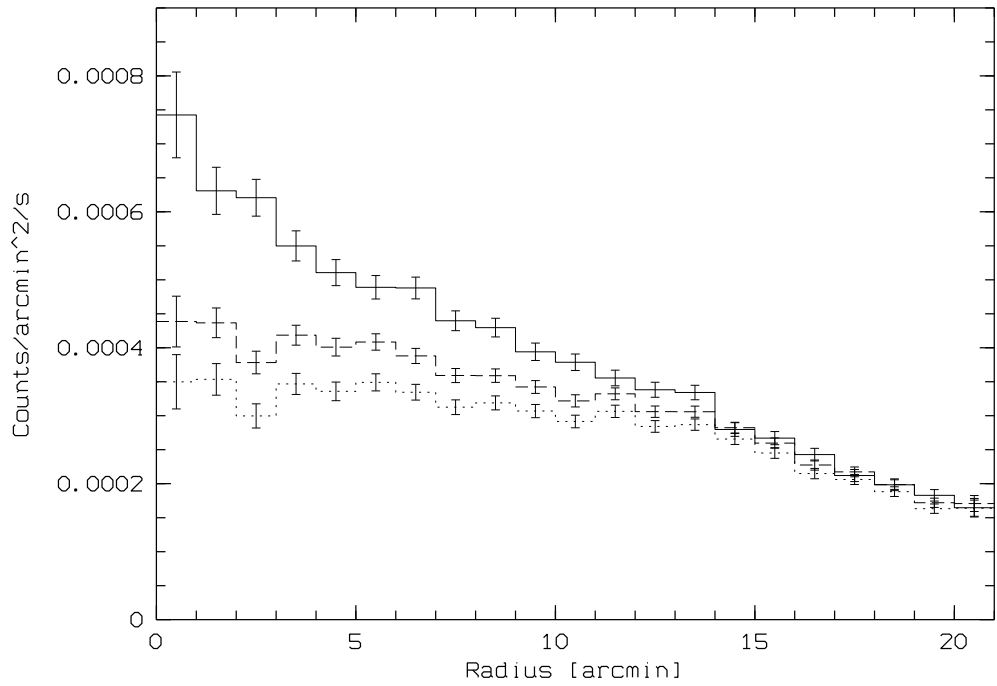,width=5.5in}
\caption{\small The azimuthally averaged radial intensity profile of the
  EUV flux in the Coma cluster is shown as a solid line. The expected EUV
  emission from the X-ray plasma is shown as a dashed line. The vignetted
  background is shown as a dotted line.}
\end{center}
\end{figure}
}

Because of these surprising results, we re-examined the previously reported
EUV excess in the Coma cluster.  We carried out our analysis using both of the
existing DS images of this cluster.  Because of the different roll orientation
and pointing position in these images, it was necessary to carry out our
analysis on each image individually.  The results were then summed and the EUV
emission and vignetted background are shown in Figure\ 9 as a solid and dotted
line respectively. In this figure, we have fit the vignetted background to the
Coma observations beyond 17\arcmin; however, because faint emission due to the
cluster probably extends past this point, especially in the direction of the NGC
4874 subcluster, this is likely to be a slight overestimate of the background
and hence the excess EUV emission we derive may be a slight underestimate. If
the X-ray profile of the Coma Cluster is used as a guide, we expect this
effect to be small compared to the statistical errors in each radial bin.

The X-ray profile has been constructed using ROSAT PSPC archival data of Coma.
We verified that our PSPC hard band cluster profile is consistent with the
profile provided by Briel, Henry \& B\"ohringer (1992) but includes the
central excess associated with the galaxy group around NGC 4874. We assumed
that this X-ray emission is due to a plasma at T = 9 keV (Donnelly et al.,
1999) absorbed by a hydrogen column of $8.7 \times 10^{19} {\rm cm}^{-2}$
(Lieu et al., 1996b) with ionization fractions and cross sections for Galactic
ISM absorption as described
above. Here we obtained a DS to PSPC hard band conversion factor of 112.
\ifpp{
\begin{figure}[tbp]
\begin{center}
\ \psfig{file=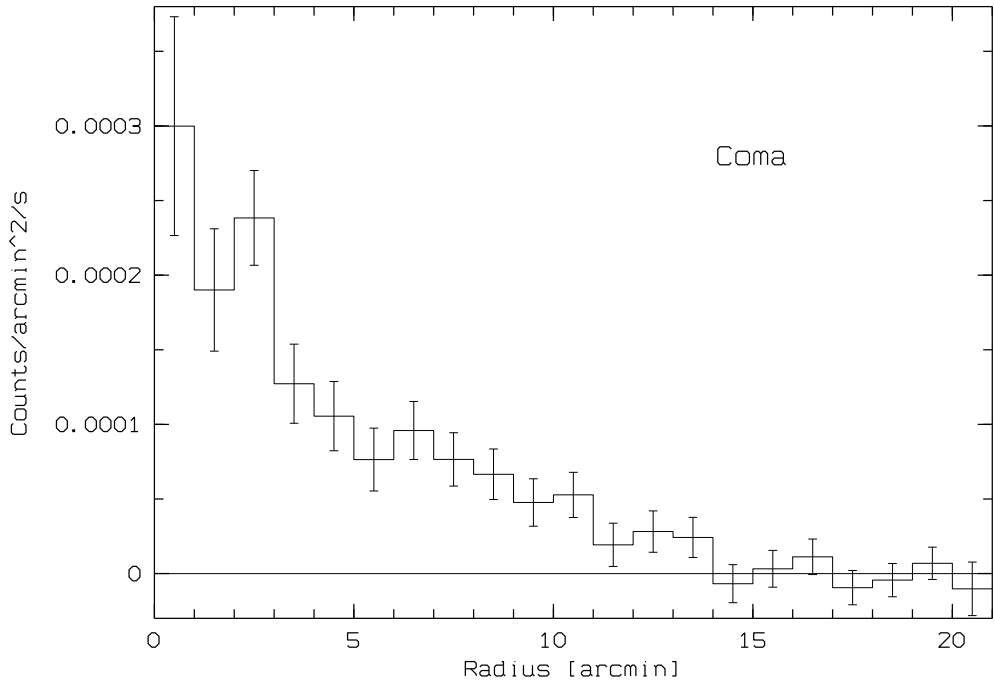,width=5.5in}
\caption{\small The excess EUV emission in the Coma cluster.}
\end{center}
\end{figure}
}

The residual EUV emission in excess of the expected contribution of the X-ray
gas, shown in Figure\ 10, demonstrates that there is, indeed, excess EUV
emission in the Coma cluster.
                        
\section{Discussion}

The results of our new analysis show no excess EUV emission at radii larger
than 4\arcmin\ for Abell 1795 (Figure\ 4) and 8\arcmin\ for Abell 2199
(Figure\ 7) contrary to previous work on these clusters. When we consider the
inner regions for these clusters, we find the results are dependent
upon a proper evaluation of the EUV emission from the X-ray plasma.  When this
emission is properly accounted for, the expected EUV emission from the
X-ray plasma, shown in Figure\ 5 for Abell 1795 and in Figure\ 8 for Abell
2199, is {\it less} than is actually produced. This can be understood in terms of excess absorption within the cluster core.
This effect has been noted in studies of X-ray cluster emission in cooling
flows, where it is often reported that the hydrogen column density is larger
in the core of the cluster. There is no observational evidence for more
hydrogen in these regions, and neutral hydrogen is not expected in this
environment. A reasonable explanation for this effect is that the X-ray
reduction codes employed in these analyses require more absorption for a
fit, and this is achieved blindly by adding more hydrogen with a
standard admixture of non-ionized metals. It is more likely that in the cooler
regions of the cooling flow, some metals are not completely ionized and these
ions produce the extra absorption of the X-ray flux (Allen et al., 1996).
This absorption would be even more substantial for the EUV flux, and would
produce the effects seen in Figures\ 5 and 8. We point out that a study of the
differing amount of absorption in the EUV and X-ray bands may provide
sufficient information to identify the primary absorbing species.

When we employ our new analysis techniques with the data on the Coma cluster
we find there {\it is} excess EUV emission in this cluster confirming the results of previous
studies. However, the distribution and intensity of this flux differs in 
detail from that previously reported.  The distribution of this radiation is
shown in Figure 10, along with the count rate intensity.  The intensity in
physical units is (slightly) dependent upon the assumed spectral distribution
of the flux.  A source with a photon spectral index of 1.6 results in an
EUV source luminosity of $1.5\pm 0.5 \times 10^{42}{\rm erg\ s^{-1}}$.

It is useful to consider why our results are different from those of Mittaz et
al. (1998). While it is difficult to evaluate the
details of another researchers' analysis, it is clear that a key difference is
our use of an observationally derived vignetted background.  Mittaz et al. and
Lieu et al.  used the theoretical background function (Richard Lieu, private communication) which is essentially
flat. In addition, these authors also carried out
their analysis of the EUV flux without first removing the non-photonic
background from their data.  The extent to which this affects the results is
not clear.  Their approach to estimating the EUV emission produced by the
X-ray plasma is also different than ours.  Given these, and perhaps
other unidentified differences, it is interesting to ask why both sets of
analyses do show excess diffuse EUV emission in the Coma cluster. The
primary explanation is that the Coma cluster does, in fact, have excess
EUV emission. This emission is sufficiently extended that the effects of 
the vignetted background, though
changing the details of the results, do not dominate as they do in Abell 1795
and Abell 2199.

We note that the Coma EUVE/soft X-ray results are confirmed by ROSAT data
despite the claim by 
Arabadjis \& Bregman (1999) that ``there is no strong evidence for
extremely soft X-ray excess in galaxy clusters.''  They reach this conclusion
using {\em only} ROSAT data and claim that an important issue that  changes the
previously published ROSAT findings is a new cross section of helium
derived by Yan et al. (1998).  
Arabadjis \& Bregman chose to parameterize the Galactic neutral hydrogen 
column in carrying out their analysis.  We note that they require more than
a $3\sigma$ deviation from measured values of this column in order to 
extinguish the Coma soft X-ray excess in the ROSAT data.  The much more
statistically robust EUVE results are not affected by the use of the Yan et al.
cross sections.

We also note that Arabadjis \& Bregman must have ``at least 50 \% of the He
in the form of \HeIII.'' This is in direct conflict with established observational results.  
Heiles et al. (1996) obtained upper limits to \HeII 268 and 269 $\alpha$ lines at
$\sim 1.4$ GHz, which rule out the possibility that any significant He III is
present in the diffuse ISM.  There is virtually no escape from this observational constraint (Heiles, private communication).
  
Our study suggests a possible reason why excess EUV emission has been found in
every cluster examined to date with EUVE. {\it Any} point in the sky will show
extended EUV emission using the analysis techniques employed in previous studies of
clusters of galaxies.

\section{Conclusions}
We obtained new data on the cluster of galaxies Abell 1795 because of concerns
that the original data set may have been compromised.  We find, however, that
these original data are valid. We investigated the effects of the telescope sensitivity variation over the field of view  and found this was a key factor in investigating extended emission. We also used a detailed approach to the evaluation of the
EUV flux produced by the X-ray gas in the core regions of this cluster. We
then expanded our study by analyzing archival data on Abell 1795, Abell 2199,
and the Coma cluster.

We find no evidence for excess EUV emission in Abell 1795 or Abell
2199. We do, however, confirm extended EUV emission in the Coma cluster
although the distribution of this flux is different in detail from that
previously reported. The fact that we do find extended EUV emission in the
Coma cluster using our new analysis procedures confirms that an unidentified
processes is operative in this cluster.

\acknowledgments

We acknowledge useful discussions with Michael Lampton, Pat Henry, John
Vallerga, Carl Heiles, Richard Lieu, and Jean Dupuis. This work
was supported in part by NASA contract NAS 5-30180. TWB was supported in part
by a Feodor-Lynen Fellowship of the Alexander-von-Humboldt-Stiftung.

\ifms{
\newpage

\figcaption[fig1.eps] { A contour plot of counts obtained in long duration DS
  exposures showing the sensitivity variation of the DS Telescope over the
  field of view. We have cut the regions at the detector ends where
  detector distortions become severe. The field displayed is approximately
  1.75 degrees x 0.73 degrees.}

\figcaption[fig2.eps] {The spatial distribution of the EUV counts in Abell
  1795.  The zero points of the image are R.\,A.$_{2000} = 13^h48^m52^s$,
  Dec.$_{2000} = +26^{\circ}35\arcmin34\arcsec$. A bright EUV emitting
  transient is visually obvious near the cluster center.  The diffuse EUV
  cluster emission peaks at the position of the central galaxy in the
  cluster.}

\figcaption[fig3.eps] {The azimuthally averaged radial intensity profile of
  the EUV flux in Abell 1795 is shown as a solid line. The vignetted
  background from long observations of blank fields is shown as a dotted line.
  There is no obvious excess EUV emission beyond 4\arcmin.}

\figcaption[fig4.eps]{The effects of different ISM effective cross sections
  and assumptions as to the ionization state on the absorption of EUV emission
  . The absorption, normalized to one at 1 keV, is shown as a function of
  energy: Baluci\'nska-Church \& McCammon (1992),dashed line; Morrison \&
  McCammon (1983) dotted line; cross section used herein (see text), solid 
  line.}

\figcaption[fig5.eps]{The expected EUV emission from the X-ray plasma in Abell
  1795 is shown as a dashed line. The azimuthally averaged radial intensity
  profile of the EUV flux is shown as a solid line. The vignetted background
  is shown as a dotted line.}

\figcaption[fig6.eps]{In 6a we show the ratio between our background
  subtracted EUV radial emission profile obtained with the EUVE Deep Survey telescope and the X-ray profile obtained with
  the ROSAT PSPC. In 6b we show the Deep Survey to PSPC count rate ratio as a function
  of radius obtained by Mittaz et al. (1998) and used by Sarazin \& Lieu
  (1998).}

\figcaption[fig7.eps]{The azimuthally averaged radial intensity profile of the
  EUV emission in Abell 2199 is shown as a solid line. The dotted line is the
  vignetted background. There is no obvious EUV emission beyond 8\arcmin.}

\figcaption[fig8.eps]{ The expected EUV emission from the X-ray plasma in
  Abell 2199 is shown as a dashed line. The EUV flux is shown as a solid line.
  The vignetted background is shown as a dotted line.}

\figcaption[fig9.eps]{The azimuthally averaged radial intensity profile of the
  EUV flux in the Coma cluster is shown as a solid line. The expected EUV
  emission from the X-ray plasma is shown as a dashed line. The vignetted
  background is shown as a dotted line.}

\figcaption[fig10.eps]{The excess EUV emission in the Coma cluster.}
}
\ifms{
\newpage
\plotone{fig1.eps}
\newpage
\plotone{fig2.eps}
\newpage
\plotone{fig3.eps}
\newpage
\plotone{fig4.eps}
\newpage
\plotone{fig5.eps}
\newpage
\plotone{fig6.eps}
\newpage
\plotone{fig7.eps}
\newpage
\plotone{fig8.eps}
\newpage
\plotone{fig9.eps}
\newpage
\plotone{fig10.eps}
}


\begin{thebibliography}{}

\bibitem[]{allen98} Allen, S., Fabian, A., Edge, A., Bautz, M., Furuzawa, A.,  
  \& Tawara, Y. 1996, \mnras, 283, 263
\bibitem[]{arbr98} Arabadjis, J. S. \& Bregman J. N. 1999, \apj, 514, 607
\bibitem[]{bamc92} Baluci\'nska-Church, M. \& McCammon, D. 1992, \apj, 400,
  699 
\bibitem[]{beet98} Bergh\"ofer, T. W., Bowyer, S., Lieu, R., \& Knude,
  J.  1998, \apj, 500, 838 
\bibitem[]{boma91}
  Bowyer, S. \& Malina, R. F. 1991, in Extreme Ultraviolet Astronomy, ed. R.
  F. Malina \& S. Bowyer (New York: Pergamon), 397 
\bibitem[]{bola96} Bowyer
  S., Lampton M., \& Lieu, R. 1996, {\em Science}, 274, 1338
\bibitem[]{boet97} Bowyer S., Lieu, R., \& Mittaz, J. P. 1998, in Proc. IAU Symp.
  188, The Hot Universe, ed. K. Koyama et al. (Dordrecht: Kluwer), 185.
\bibitem[]{bobe98} Bowyer,S. \&
  Bergh\"ofer, T. W. 1998, \apj, 506, 502 
\bibitem[]{brhebo92} Briel, U. G., Henry, J. P. \& B\"ohringer, H. 1992, \aap,
  259L, 31
\bibitem[]{brhe96}
  Briel, U. \& Henry, J. P. 1996, \apj, 472, 131 
\bibitem[]{ceno98} Cen, R. \& Ostriker, J. 1999, \apj, in press 
\bibitem[]{donn99} Donnelly, R. H., Markevitch, M., Forman, W., Jones, C.,
  Churazov, E., \& Gilfanov, M. 1999, \apj, in press 
\bibitem[]{enbi98} En{\ss}lin, T. \& Biermann, P. 1998, \aap, 330, 96
\bibitem[]{enet98} En{\ss}lin, T., Lieu, R., \& Biermann, P. 1999, \aap,
 in press 
\bibitem[]{fuet98} Fusco-Femiano, R., Dal Fiume, D., Feretti, L., Giovannini, G, Matt, G., Molendi, S., \& Santangelo, A. 1999,
  \apj, 513, L21 
\bibitem[]{hklr96} Heiles, C., Koo, B.~C., Levenson, N., \& Reach, W. 1996, \apj, 462, 326
\bibitem[]{hw97} Hwang, C.-Y. 1997, {\em Science}, 278, 1917 
\bibitem[]{ki72} King, I. 1972, \apj, 174, L123 
\bibitem[]{kobo98} Korpela, E.,\& Bowyer S. 1998, \apj, 115, 2551 
\bibitem[]{limi96} Lieu, R., Mittaz, J., Bowyer ,S., Lockman, F., Hwang, C.Y., 
\& Schmitt, Y. 1996a, \apjl, 458,
  L5
\bibitem[]{limib96} Lieu, R., Mittaz, J., Bowyer, S., Breen, J., Lockman, F., 
Murphy, E., \& Hwang, C.Y. 1996b, Science, 274, 1335 
\bibitem[]{liip98} Lieu, R., Ip, W-H., Axford, W., \& Bonamente, M. l999a,
  \apjl, 510 , L25 
\bibitem[]{lieu99} Lieu, R., Bonamente, M., \& Mittaz, J. 1999b, \apj, 517, L91
\bibitem[]{mili98} Mittaz, J., Lieu, R., \& Lockman, F. l998, \apjl, 498, L17
\bibitem[]{momc83} Morrison, R. \& McCammon, D. 1983, \apj, 270, 119
\bibitem[]{rees88} Rees, M. J. 1988, \nat, 333, 523 
\bibitem[]{reph99} Rephaeli, Y., Gruber, D., \& Blanco, P. 1999,
\apj, 511, 21
\bibitem[]{rey98} Reynolds, R. J., Tufte, S. L., Haffner, L. M., Jaehnig, K., 
\& Percival, J.~W. 1998, Pub. Astron. Soc. Aus., 15, 14
\bibitem[]{rubo94} Rumph, T., Bowyer, S., \& Vennes, S. 1994, AJ, 107, 2108 
\bibitem[]{sali98} Sarazin, C. \& Lieu, R. l998, \apj, 494, L177
\bibitem[]{sidd98} Siddiqui, H., Stewart, G., \& Johnstone, R. 1998, 
\aap, 334, 71
\bibitem[]{siet97} Sirk, M. M., Vallerga, J. V., Finley, D. S., Jelinsky, P., \&
Malina, R. 1997,
  \apjs, 110, 347 
\bibitem[]{varo97} Vallerga, J. \& Roberts, B. 1997, CEA memo: EUVE/20/97 
\bibitem[]{yan}Yan, M., Sadeghpour, H., \& Dalgarno, A. 1998, \apj, 496, 1044
\end{thebibliography}
\end{document}